\documentclass[a4paper,11pt]{article}
\usepackage{jinstpub} %
\usepackage{lineno}

\usepackage{siunitx}
\usepackage{booktabs}
\usepackage{xspace}
\usepackage{subcaption}
\usepackage{graphicx}
\usepackage{color}
\usepackage{bm}
\usepackage{soul}
\usepackage{float}

\newcommand{\geant}{\textsc{Geant4}\xspace}
\newcommand{\caloclouds}{\textsc{CaloClouds}\xspace}
\newcommand{\ccedm}{\textsc{CaloClouds~II}\xspace}
\newcommand{\cccm}{\textsc{CaloClouds~II~(CM)}\xspace}

\definecolor{airforceblue}{rgb}{0.36, 0.54, 0.66}
\definecolor{darkpurple}{rgb}{0.5, 0.2, 0.8}
\definecolor{dartmouthgreen}{rgb}{0.05, 0.5, 0.06}
\definecolor{taborange}{rgb}{1.0, 0.49, 0.054}
\definecolor{amaranth}{rgb}{0.9, 0.17, 0.31}

\title{CaloClouds II: Ultra-Fast Geometry-Independent Highly-Granular Calorimeter Simulation}

\author[a]{Erik Buhmann,}
\author[b,c]{Frank Gaede,}
\author[a,c]{Gregor Kasieczka,}
\author[b,1]{Anatolii Korol,\note{Corresponding author.}}
\author[a]{William Korcari,}
\author[b]{Katja Kr\"uger}
\author[b]{and Peter McKeown}

\affiliation[a]{Institut f\"ur Experimentalphysik, Universit\"at Hamburg,\\Luruper Chaussee~149, 22607 Hamburg, Germany}
\affiliation[b]{Deutsches Elektronen-Synchrotron DESY,\\Notkestr. 85, 22607 Hamburg, Germany} 
\affiliation[c]{Center for Data and Computing in Natural Sciences CDCS, Deutsches Elektronen-Synchrotron DESY,\\Notkestr. 85, 22607 Hamburg, Germany}

\emailAdd{anatolii.korol@desy.de}

\abstract{
Fast simulation of the energy depositions in high-granular detectors is needed for future collider experiments at ever-increasing luminosities. 
Generative machine learning (ML) models have been shown to speed up and augment the traditional simulation chain in physics analysis. 
However, the majority of previous efforts were limited to models relying on fixed, regular detector readout geometries.
A major advancement is the recently introduced \caloclouds model, a geometry-independent diffusion model, which generates calorimeter showers as point clouds for the electromagnetic calorimeter of the envisioned International Large Detector (ILD).

In this work, we introduce \ccedm which features a number of key improvements. This includes continuous time score-based modelling, which allows for a 25-step sampling with comparable fidelity to \caloclouds while yielding a $6\times$ speed-up over \geant on a single CPU ($5\times$ over \caloclouds).
We further distill the diffusion model into a consistency model allowing for accurate sampling in a single step and resulting in a $46\times$ speed-up over \geant ($37\times$ over \caloclouds). This constitutes the first application of consistency distillation for the generation of calorimeter showers. 
}


\arxivnumber{2309.05704} %

\begin{document}
\maketitle
\flushbottom

\section{Introduction}
\label{sec:Intro}

Accurate simulations of particle physics experiments are crucial for comparing theory predictions with experimental results.
With the planned high luminosity upgrade to the Large Hadron Collider (LHC)~\cite{HLLHC} and other envisioned collider experiments like those at the International Linear Collider (ILC)~\cite{ILC-TDR}, experimental data is going to be taken at ever increasing rates.
The amount of simulated events needs to keep up with these rates, which is difficult to achieve with current Monte Carlo simulations and the projected computing budgets at large experiments~\cite{HEPSoftwareFoundation:2017ggl,LHCC_HL:2022}.

Detector simulations, such as the simulation of the sensor response in highly granular calorimeters, can be augmented or sped up by employing modern generative machine learning methods~\cite{CaloGan1,ganplify, Calomplification,Adelmann:2022ozp}. 
Recent studies have explored the simulation of calorimeter showers with various generative models such as generative adversarial networks (GANs)~\cite{CaloGan1, CaloGan2, CaloGan3, ErdmannWGAN1, ErdmannWGAN2, Belayneh_2020_calo_w_DL, ATL-SOFT-PUB-2020-006_ATLAS_fastsim_GAN, Vallecorsa2,  DetectorSim2, AtlasCaloFastsim, AtlFast3, Hashemi:2023ruu}, autoencoders and their variants~\cite{gettinghigh, decoding_photons, hadrons, atlascollaboration2022deep_generative_models_fast_photon_showers, cresswell2022caloman, Diefenbacher:2023angles}, and normalizing flows~\cite{fast_sim_at_LHC_autoregressive2021, krause2021caloflow, krause2021caloflow2,schnake2022_pointFlow,krause2023caloflow_forCaloChallenge_DS1, Diefenbacher:2023prl, xu2023generative_detector_response_wiht_Cond_NomalizingFlows, buckley2023inductive_caloFlow}.
Additionally, diffusion models~\cite{sohldickstein2015deep, song2020generative_estimatingGradients, song2020improved_technieques_for_sorebased_geneerative, ho2020denoising, song2021scorebased_generativemodelling}, also referred to as score-based generative models,
have been shown to provide very high fidelity on calorimeter data~\cite{Mikuni:2022xry_CaloScore, CaloClouds, acosta2023comparison_imageVSpointClouds, mikuni2023caloscore_v2, amram2023calodiffusion}.
Beyond detector simulation, generative models have, for example, also been explored as event generators~\cite{EventGen1, EventGen2, EventGen3, EventGen4, EventGen5, EventGen6, EventGen7} and parton shower simulators~\cite{de_Oliveira_2017_learning_particle_physics_by_example, Andreassen_2019_JUNIPR, Bothmann_2019_ReweightingPartonShower, HadronizationGen2, kansal2022particle, kaech2022jetflow, kaech2022point_transformer_refiner, Kansal_2023, Buhmann:2023pmh, leigh2023pcjedi, kaech2023attention_JetNet_MDMA, butter2023jet_diffusion, leigh2023pcdroid}.

Most previous generative calorimeter models rely on a fixed data geometry, representing calorimeter showers as 3-dimensional images with the energy as the ``color'' channel and each pixel representing a calorimeter sensor. 
Modern high granularity calorimeters consist of many thousands of sensor cells or more (e.g. 6 million for the planned CMS HGCal~\cite{CMS_hgcal}), 
but a given shower often deposits energy in only a small fraction of cells resulting in very sparse 3d image representations.
Hence, it is much more computationally efficient to only simulate the actual energy depositions with a generative model. This can be achieved by describing the shower with only the coordinates and energies deposited --- i.e.  a \textit{point cloud}. 
Such a multidimensional calorimeter point cloud can be represented by four features, the three-dimensional spatial coordinates and the cell energy, with the number of points equivalent to the number of cells containing hits.

In addition to computational efficiency, 
such point cloud showers have the major advantage that they can represent not only cell energies, but also much more granular \geant step information, i.e. simulated energy depositions in the material, not accessible in experiments.
Such \geant step point clouds are largely independent of the cell structure within a layer of a given calorimeter, effectively allowing the translation-invariant projection of the shower into any part of the calorimeter, regardless of cell type.
These projections with \geant step point clouds are less likely to produce artifacts due to gaps or cell staggering than cell-level point clouds would, resulting in a largely geometry-independent description of the calorimeter shower.
This approach is complimentary to a geometry-aware model~\cite{Liu_2023_geometry_aware}, which is trained with a dataset containing various calorimeter geometries.

Previous point cloud and graph generative models explored in particle physics~\cite{kansal2022particle, Kansal_2023, Buhmann:2023pmh, leigh2023pcjedi, mikuni2023fast_pointcloudDiffussion, kaech2023attention_JetNet_MDMA, leigh2023pcdroid} were only used for relatively small numbers of points. However, energetic calorimeter showers in high granularity calorimeters consist of $\mathcal{O}(1000)$ points. 
To generate such showers, we recently introduced \caloclouds~\cite{CaloClouds} a generative model able to accurately generate photon showers in the form of point clouds with several thousands of points (namely clustered \geant steps), in order to achieve geometry-independence. 
Since then, a specific comparison between a generative model for fixed geometry and a generative model for point cloud structured calorimeter showers on cell-level was performed in Ref.~\cite{acosta2023comparison_imageVSpointClouds}.

This \caloclouds architecture consists of multiple sub-models with a diffusion  model (see Sec.~\ref{sec:Diffusion_Model} for details) at its core.
Most diffusion models, including the one used in \caloclouds, are currently held back by their slow sampling speed, as many evaluation steps have to be performed to generate events. 
However, recent advances in computer vision achieve very high generative fidelity on natural image data with $\mathcal{O}(10)$ model evaluations using advanced training paradigms and novel ordinary and stochastic differential equation 
solvers~\cite{zhang2023fast_sampling_diffusion_models, karras2022elucidating_EDM_diffusion, lu2022dpmsolver, lu2023dpmsolver_++}. 
In this work, we first leverage recent advances in the training and sampling procedure of diffusion models in order to generate samples with the \ccedm model\footnote{The code is available at \url{https://github.com/FLC-QU-hep/CaloClouds-2}.} using much fewer model evaluations than the original \caloclouds model, %
by following the diffusion paradigm introduced in Ref.~\cite{karras2022elucidating_EDM_diffusion}.%

Another research direction to speed up  generative models is the distillation of diffusion models into models which require significantly fewer function evaluations during sampling than the original model~\cite{luhman2021knowledge_distillation, salimans2022progressive, liu2022flow_matching_distllation, zheng2023fast_sampling_diffusion_distllation, berthelot2023tract_distillation}.
Recently, consistency models have been introduced as a novel kind of generative model allowing for single and multi-step data generation~\cite{song2023consistency}. 
These consistency models can either be trained ab-initio or distilled from an already trained diffusion model. 
We demonstrate the ability to distill our diffusion model into a consistency model, 
thereby allowing data generation with a single model evaluation leading to further speed-ups.

\pagebreak
In summary, the proposed \ccedm contains the following adjustments:
\begin{enumerate}
    \item The previously used 
    discrete-time diffusion process is replaced with the continuous-time diffusion paradigm introduced in Ref.~\cite{karras2022elucidating_EDM_diffusion}. This allows for fewer diffusion iterations during sampling. 
    \item The common latent space is removed as we have noticed no advantage for the generative fidelity when generating photon calorimeter showers. This removal yields a simplified model architecture and improved training and sampling speeds.
    \item We add a calibration to the energy per calorimeter layer as well as applying a calibration to the center of gravity in the $X$- and $Y$-direction of the generated point cloud showers. This replaces the previous total energy calibration and improves the generative fidelity in the longitudinal energy distribution. 
    \item Further, we apply consistency distillation to distill the diffusion model into a consistency model~\cite{song2023consistency}, allowing single step generation and therefore greatly improved sampling speed. We refer to this model as \cccm. 
\end{enumerate}

In  Sec.~\ref{sec:Data} we describe the point cloud dataset used for training and evaluation.
The diffusion paradigm and model components of the \ccedm model are explained in Sec.~\ref{sec:Model}.
We compare the generative fidelity of \ccedm and its variant to the original \caloclouds model in Sec.~\ref{sec:Results} and draw our conclusions in Sec.~\ref{sec:Conclusion}.

\section{Data Samples}
\label{sec:Data}

To compare the performance of our improved \ccedm model we use the same dataset as in Ref.~\cite{CaloClouds}.
The data describes a calorimeter shower in the form of a point cloud.
Each calorimeter shower consists of energy depositions of photons showering in a section of the high-granular electromagnetic calorimeter (ECAL) of the envisioned International Large Detector (ILD)~\cite{ILD-IDR}.
As a sampling calorimeter, it consists of 30 layers with passive tungsten material and active silicon sensors.
All individual silicon layers consist of small $5 \times 5\,\text{mm}$ readout cells with a thickness of $0.5\,\text{mm}$. 
Between the first 20 active layers in the longitudinal direction there are passive layers with a thickness of $2.1\,\text{mm}$ and between the remaining 10 layers the passive layers have a thickness of $4.2\,\text{mm}$.
We simulated the dataset with \geant Version 10.4 (using the QGSP\_BERT physics list) implemented in the iLCSoft framework~\cite{ilcsoft}. The simulated geometric model is implemented in \textsc{DD4hep}~\cite{dd4hep} and includes realistic gaps between the sensors and position dependent irregularities. 
More simulation details can be found in Ref.~\cite{CaloClouds}.

During the full \geant simulation up to 40,000 individual energy depositions 
originating from secondary particles traversing the active sensor material
are registered (depending on the incident photon energy). 
These energy depositions are commonly referred to as \geant \textit{steps}.
All steps that fall into the volume of the same sensor are subsequently summed, resulting in the energy deposited in a cell \textit{hit}. 
These cell hits (up to 1,500 at $90~\text{GeV}$) are then used for downstream analysis as it is the same low-level information that is measurable in a real experimental setting.

Ideally, a generative model should produce cell-level hits to make the full \geant simulation more computationally efficient.
Cell-level information is also generated in all other approaches for fast calorimeter shower simulation with generative machine learning models.
However, generating discrete cell hits directly in the form of a point cloud is challenging, as minor imperfections such as generating multiple points in the same calorimeter cell can heavily impact the generative fidelity in various high level observables like the total number of cell hits $N_{hits}$. 

Therefore, it could be advantageous to generate point clouds not on hit-level but on \geant step-level, i.e. many simulated very granular energy depositions per cell, resulting in much larger point clouds where points are 
continuously distributed in space (as opposed to discrete cell hits).
Yet, we found generating a point cloud with up to 40,000 steps prohibitively expensive %
from a computational point-of-view. Additionally such a high resolution is not necessary for good generative fidelity. 
Therefore in Ref.~\cite{CaloClouds} we introduced a middle ground: we cluster the up to 40,000 \geant steps into up to 6,000 \textit{points}. 
For this clustering, the steps are grouped into their layer and their energy is binned in an ultra-high granularity grid with $36\times$ higher granularity than the cell resolution, resulting in a square grid %
size of $0.83 \times 0.83\,\text{mm}^2$. 
This results in a clustered point cloud of up to 6,000 points --- sufficiently small to be generated with the \caloclouds model, yet distributed in discrete positions with sufficiently small separation so as to be approximately a continuous point distribution in 3d space. 

In addition to a computationally efficient simulation, this makes the generated calorimeter point cloud largely geometry-independent of the actual cell layout of the calorimeter, unlike point clouds based on cell-level energy depositions. 
This ultra-high granularity calorimeter point cloud can be projected into any part of the calorimeter (except changing its depth), without introducing reconstruction artifacts due to for example gaps and cell staggering, as successfully shown in Ref.~\cite{CaloClouds}.

To produce the training set, a total of 524,000 showers were generated with \geant, with an incident energy uniformly sampled between $10$ and $90~\text{GeV}$. 
Additionally, multiple test sets were generated: 40,000 showers uniformly distributed in energy for the figures shown in Sec.~\ref{sec:Results_Physics}; 2,000 showers for the single energy plots at $10$, $50$, and $90~\text{GeV}$; and 500,000 showers for calculating the evaluation metrics and the classifier score in Sec.~\ref{sec:Results_Evaluation} and Sec.~\ref{sec:Results_Classifier}. 

Each point of the point cloud has four features: the $X$- and $Y$-position (transverse to the incident particle direction), the $Z$-position (parallel 
to the incident particle direction), and the $energy$.
As a pre-processing step, the passive material regions are removed such that the point locations in the dataset also become continuous in the longitudinal $Z$-axis.
The position features, $X$, $Y$, and $Z$, are each %
normalized to the range $[-1,1]$.
The energy feature of the 4d point cloud is given in MeV.

As it is important for downstream analyses to accurately simulate the behaviour of photon showers on 
the level of the physical geometry, i.e. at cell level, all results shown in Sec.~\ref{sec:Results_Physics} to \ref{sec:Results_Classifier} are on cell-level.
To this end, the calorimeter point cloud --- with either up to 40,000 points for \geant or with up to 6,000 points for those generated with \caloclouds / \ccedm --- are binned to the realistic ILD ECAL layout (including detector irregularities and gaps) with $30 \times 30 \times 30$ calorimeter cells.

\section{Generative Model}
\label{sec:Model}

The \ccedm model is an improved version of the original \caloclouds architecture from Ref.~\cite{CaloClouds}.
First, we revisit the main model components of the \caloclouds model, afterwards we outline the improvements made in \ccedm.

\caloclouds is a combination of two normalizing flows~\cite{AutoregressivFLow}, a VAE-like encoder~\cite{kingma_welling_variational_autoencoder}, and a discrete time Denoising Diffusion Probabilistic Model (DDPM)~\cite{ho2020denoising}. 
Specifically, it consists of the \textit{Shower Flow}, a normalizing flow generating conditioning and calibration features;  the \textit{EPiC Encoder}, an encoder based on Equivariant Point Cloud (EPiC) layers~\cite{Buhmann:2023pmh} to encode calorimeter showers during training into a latent space for model conditioning; the \textit{Latent Flow}, a normalizing flow trained to model the encoded latent space during sampling; and a diffusion model, called \textit{PointWise Net}, which is a DDPM-based diffusion model generating each point independent and identically distributed  (i.i.d.)\ based on a common latent space, incident energy and number of points conditioning. 
The models are implemented using \textsc{PyTorch 1.13}~\cite{pytorch}. 

In the following, we outline the differences between \caloclouds and \ccedm. 
The largest conceptual difference is the change of the diffusion paradigm. We move from a discrete time diffusion process (DDPM), in which the training and sampling is performed with the same number of diffusion steps, to a continuous time diffusion paradigm based on Ref~\cite{karras2022elucidating_EDM_diffusion}, sometimes referred to as \textit{EDM} diffusion or \textit{k-diffusion}. 
This EDM diffusion allows for training a continuous time score function, which can be used to denoise any noise level,
thereby separating the training and sampling procedure and allowing for sampling with various ordinary differential equation (ODE) and stochastic differential equation (SDE) solvers and different step sizes. %
Crucially, it allows to trade off sampling speed and sampling fidelity without retraining.
We find good performance with the 2$^\text{nd}$-order Heun ODE solver and the step size parameterisation suggested in Ref~\cite{karras2022elucidating_EDM_diffusion}. 
Additional details on the diffusion paradigm is given in the following Sec.~\ref{sec:Diffusion_Model}.
 
As a second change, \ccedm simplifies the original model.
We noticed that for the photon calorimeter shower point clouds we are generating in this study, the shared latent space between points is not necessary for high generative fidelity. Therefore the latent dimensionality can be set to zero, so the \textit{EPiC Encoder} and the \textit{Latent Flow} are removed.
By discarding it we achieve a simpler model as well as improved training and sampling efficiency. 

Next, the \textit{Shower Flow} for generating conditioning and calibration features is expanded to generate the total number of points, total visible energy, the relative number of points and energy of each calorimeter layer in the $Z$-direction, as well as the center of gravity in the $X$- and $Y$-direction. 
This flow is conditioned on the incident particle energy only.
The total number of points generated per shower is used --- together with the incident particle energy --- for the conditioning of the PointWise Net diffusion model.

Overall, the Shower Flow is composed of ten blocks, each %
with seven coupling layers~\cite{RealNVP, durkan2019neural} conditioned on the incident particle energy. 
It is implemented using the \textsc{Pyro} package~\cite{bingham2019pyro}. 
The Shower Flow is trained once for 350k iterations and used for all three models (\caloclouds, \ccedm, and \cccm) compared in Sec.~\ref{sec:Results}.

The post-diffusion calibration expands upon the calibration in Ref.~\cite{CaloClouds}:
The number of hits per layer is calibrated by ordering all points in the $Z$-coordinate and setting iteratively the first $N_{z,i=1}$ points to $z_i = 1$ (first layer), the second $N_{z,i=2}$ points to $z_i = 2$ (second layer) and so on until the $30^{th}$ layer.
Afterwards, we calibrate the total layer energy by re-weighting each point energy to sum up to the predicted layer energy $E_{\text{pred},i}$.
Finally, we calculate the center of gravity in $X$ and $Y-$direction of the point cloud and subtract its difference in comparison to the predicted center of gravity from each point's $X-$ and $Y-$ coordinate to calibrate the overall point cloud center of gravity in these two dimensions.
Note that we further set points with negative generated energy to zero.

\begin{figure}[tbp]
    \begin{subfigure}[h]{0.40\textwidth}
         \centering
         \includegraphics[width=\textwidth]{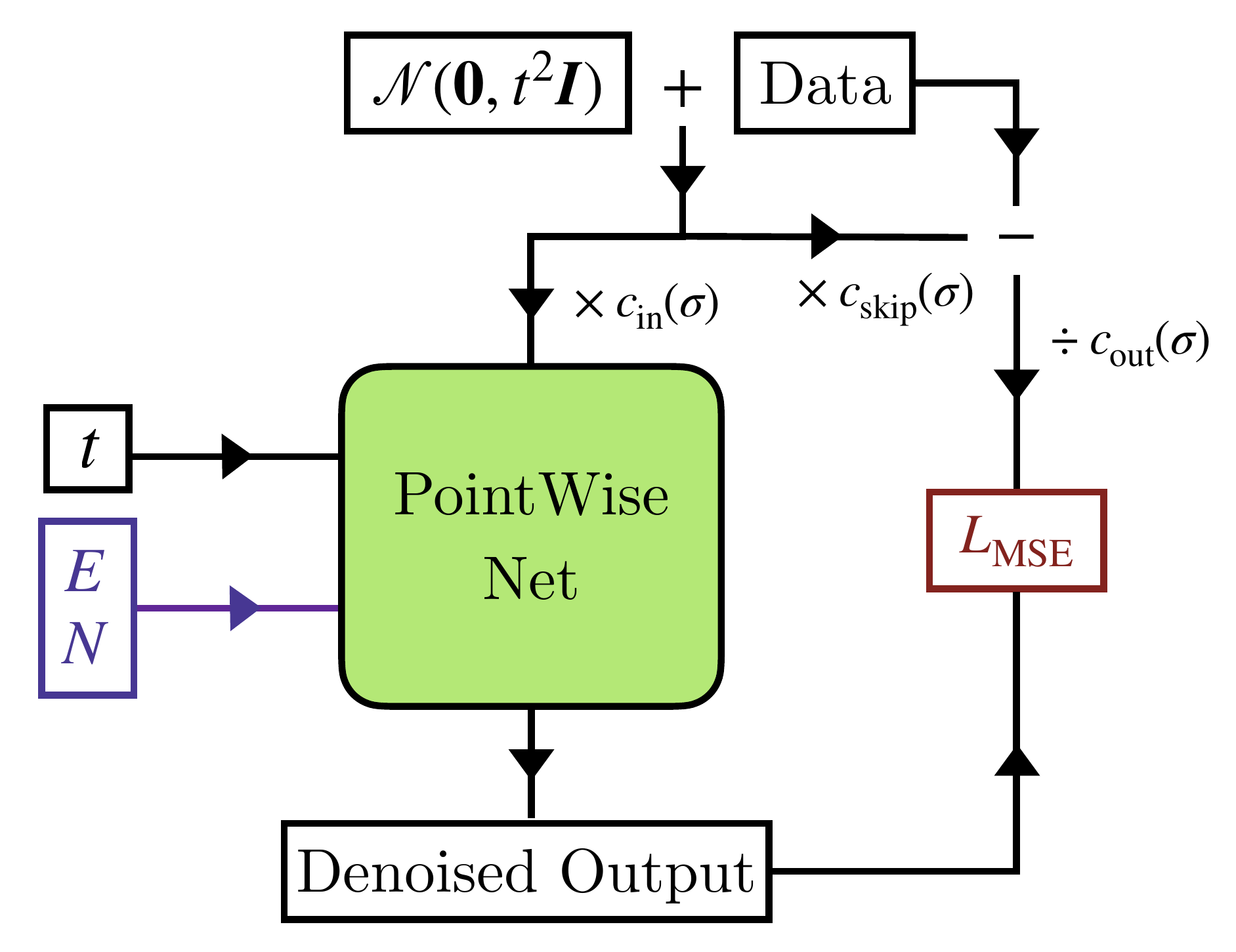}
         \caption{Training}
         \label{fig:training}
    \end{subfigure}
    \hspace{0.1pt}
    \begin{subfigure}[h]{0.55\textwidth}
         \centering
         \includegraphics[width=\textwidth]{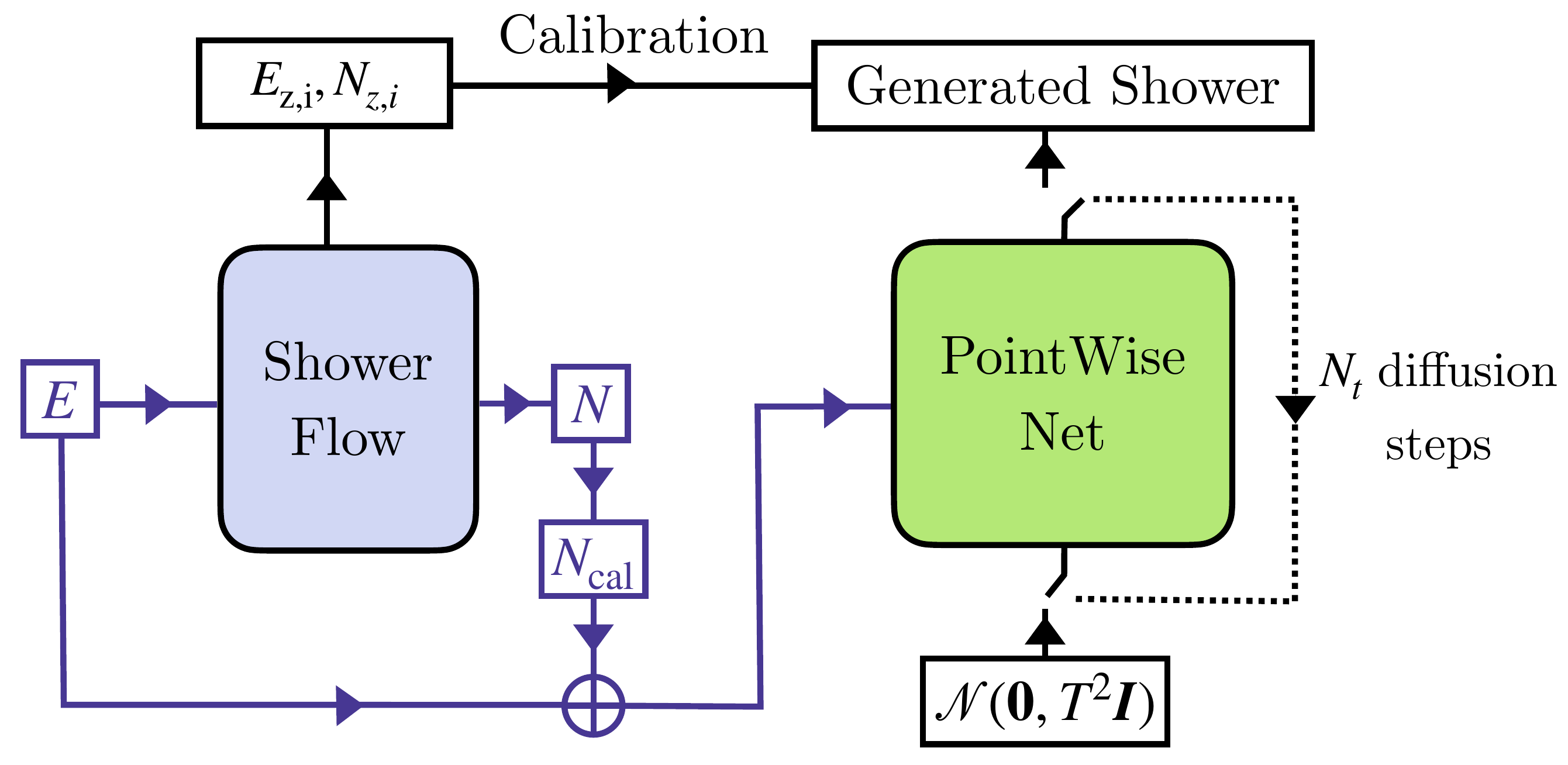}
         \caption{Sampling}
         \label{fig:sampling}
    \end{subfigure}
\caption{
Illustration of the training and sampling procedure of the \ccedm model.
\textbf{(a)} During training a random continuous time step $t$ is trained conditioned on the shower energy $E$ and number of points $N$. The loss, $L_\text{MSE}$, is approximated by a simple mean squared error (MSE) between the noised data and the denoised output. 
The scaling functions $c_\text{in}$, $c_\text{out}$, and $c_\text{skip}$ are defined following Eq.~\ref{eq:scaling}.
\textbf{(b)} During sampling the $E$-conditional Shower Flow generates $N$ as well as shower observables for calibration. After a $N$ calibration the PointWise Net denoises iteratively noise $\mathcal{N}(\boldsymbol{0}, T^2 \boldsymbol{I})$ into a calorimeter shower. 
When sampling with \cccm only one denoising step is performed.
}
\label{fig:training_sampling_diagram}
\end{figure}

During sampling, the  number of points predicted by the Shower Flow is calibrated before being used for the conditioning of the Latent Flow and the PointWise Net. 
The calibrated number of points is given by $N_\mathrm{cal} = N_\mathrm{uncal} \cdot p_\mathrm{gen}(p_\mathrm{data} (N_\mathrm{uncal}))$, where $p_\mathrm{data}$ is a cubic polynomial fit of the ratio of 
the number of points $N_\mathrm{uncal,\,G4}$ to the number of cell hits $N_\mathrm{cell,\,G4}$ of the \geant showers and $p_\mathrm{gen}$ is a fit of the ratio of 
number of cell hits $N_\mathrm{cell,\,gen}$ to the (uncalibrated) number of points $N_\mathrm{uncal,\,gen}$ of a given model.
Hence, this polynomial fit $p_\mathrm{gen}$ is performed for each model separately. 
More details on the model components and the calibrations can be found in Ref~\cite{CaloClouds}.
A schematic overview of the training and sampling procedure is shown in Fig.~\ref{fig:training_sampling_diagram}. 

In the following Sec.~\ref{sec:Diffusion_Model} we describe the continuous time diffusion paradigm implemented in the \ccedm model and in Sec.~\ref{sec:Consistency_Model} we outline its distillation into a consistency model, referred to as \cccm.
Both models use the same model components outlined above.
Details on the training and sampling hyperparameters %
are outlined in Sec.~\ref{sec:training_and_sampling}.

\subsection{Diffusion Model}
\label{sec:Diffusion_Model}

The diffusion model~\cite{sohldickstein2015deep} used in the \caloclouds model is a Denoising Diffusion Probabilistic Model (DDPM) with the same discrete time steps during model training and sampling~\cite{ho2020denoising, luo2021diffusion}. 
Since the introduction of DDPM, subsequent works, i.e. Refs.~\cite{song2021scorebased_generativemodelling, song2022denoising_DDIM, karras2022elucidating_EDM_diffusion}, have shown that it is advantageous to train a diffusion model with continuous time conditioning. 
This allows for a more flexible sampling regime for which various SDE and ODE solvers with either a fixed or an adaptive number of solving steps can be applied. 

In the following, we outline the key parts of a diffusion model based on the paradigm outlined in Ref.~\cite{karras2022elucidating_EDM_diffusion}.
The training of a diffusion model starts by diffusing a data distribution $p_\mathrm{data}(x)$ with an SDE~\cite{song2021scorebased_generativemodelling} in the forward direction (``data'' $\rightarrow$ ``noise'') via
\begin{equation}
    \mathrm{d} \mathbf{x}_t=\boldsymbol{\mu}\left(\mathbf{x}_t, t\right) \mathrm{d} t+\sigma(t) \mathrm{d} \mathbf{w}_t  ,
    \label{eq:sde}
\end{equation}
where $t$ is a fixed time step defined in the interval $t \in [0,T]$ with $T>0$ as a hyperparameter. 
$\boldsymbol{\mu}(\cdot, \cdot)$ and $\sigma(\cdot)$ denote the \textit{drift} and \textit{diffusion} coefficients, and $  \mathbf{w}_{t \in [0,T]}$ is the standard Brownian motion.  
The distribution of $\mathbf{x}_t \sim p_t(\mathbf{x}) = p_\text{data} (\mathbf{x}) \ast \mathcal{N}(\mathbf{0}, T^2 \boldsymbol{I})$ ($\ast$ as the convolution operator) and at time step zero it is identical to the data distribution $p_0(\mathbf{x}) = p_\mathrm{data}(\mathbf{x})$.
When reversing this diffusion process (``noise'' $\rightarrow$ ``data''), a so called \textit{probability flow ODE} emerges with a solution trajectory sampled at time step $t$ given by
\begin{equation}
    \mathrm{d} \mathbf{x}_t=\left[\boldsymbol{\mu}\left(\mathbf{x}_t, t\right)-\frac{1}{2} \sigma(t)^2 \nabla \log p_t\left(\mathbf{x}_t\right)\right] \mathrm{d} t ,
    \label{eq:PF_ODE}
\end{equation}
with $\nabla \log p_t\left(\mathbf{x}_t\right)$ as the \textit{score function} of $p_t(\mathbf{x})$.
As suggested in Ref.~\cite{karras2022elucidating_EDM_diffusion}, we set the coefficients in the SDE in Eq.~\ref{eq:sde} to $\boldsymbol{\mu}(\mathbf{x},t) = 0$ and $\sigma(t) = \sqrt{2t}$ to ensure that $p_T(\mathbf{x})$ is close to the distribution $\mathcal{N}(\mathbf{0}, T^2 \boldsymbol{I})$.
Since the exact analytical score function is usually unknown, we train a neural network with weights $\phi$ as a score model $\boldsymbol{s}_\phi(\mathbf{x},t) \approx \nabla \log p_t\left(\mathbf{x}_t\right)$ to get the empirical probability flow ODE:
\begin{equation}
    \frac{\mathrm{d} \mathbf{x}_t}{\mathrm{~d} t}=-t \boldsymbol{s}_\phi\left(\mathbf{x}_t, t\right)
    \label{eq:empirical_PF_ODE}
\end{equation}

For the purpose of numerically stable scaling behaviour, we follow Ref.~\cite{karras2022elucidating_EDM_diffusion} and actually train a separate network $\boldsymbol{d}_\phi$ with $t$-dependent skip connections from which $\boldsymbol{s}_\phi$ is derived:
\begin{equation}
    \boldsymbol{s}_\theta(\mathbf{x} , t)=c_{\text{skip}}(t)\ \mathbf{x}+c_{\text{out}}(t)\  \boldsymbol{d}_\theta\left(c_{\text{in}}(t)\ \mathbf{x} , t\right)
    \label{eq:scaling}
\end{equation}
The coefficients are time dependent and control the skip connection via $c_\text{skip} = \sigma_\text{data}^2 / \left(\sigma_\text{data}^2 + t^2 \right)$, 
the input scaling via $c_\text{in} = t \cdot \sigma_\textbf{data} / \sqrt{\sigma_\text{data}^2 + t^2}$ 
and the output scaling via $c_\text{out} = 1 / \sqrt{\sigma_\text{data}^2 + t^2}$.
The hyperparameter $\sigma_\text{data}$ corresponds roughly to the variance of $p_\text{data}(\mathbf{x})$ and is set to $\sigma_\text{data} = 0.5$. 
During training a random time step is drawn from the continuous noise distribution $\text{ln}(t) = \mathcal{N}\left( P_\text{mean}, P_\text{std}^2 \right)$, with $P_\text{mean} = -1.2$ and $P_\text{std} = 1.2$ (the default parameters chosen in Ref.~\cite{karras2022elucidating_EDM_diffusion}), and the loss is given by:
\begin{equation}
    \mathbb{E}_{t,\mathbf{x}_t,\mathbf{x}_0} \left[ \| \boldsymbol{s}_\theta(\mathbf{x}_t , t) - \mathbf{x}_0 \|_2^2 \right]
\end{equation}
An illustration of this training process can be found in Fig.~\ref{fig:training}.

For sampling from the trained score model, one samples from noise at time step $T$ as $\mathbf{\hat{x}}_T \sim \mathcal{N}(\mathbf{0}, T^2 \boldsymbol{I})$ and integrates the probability flow ODE in Eq.~\ref{eq:PF_ODE} 
over discrete time steps backwards in time using a numerical ODE solver. 
This results in a sample $\mathbf{\hat{x}}_0$ which provides a good approximation of a sample from the data distribution $p_\mathrm{data}(\mathbf{x})$.
In practice the solver is usually stopped at a small positive value $\epsilon > 0$ to avoid numerical instabilities resulting in the approximate sample $\mathbf{\hat{x}}_\epsilon \approx \mathbf{\hat{x}}_0$.
For our sampling, we use the suggested values and step scheduling from Ref.~\cite{karras2022elucidating_EDM_diffusion} with $T = 80$ and $\epsilon = 0.002$, and apply the $2^\text{nd}$ order Heun ODE solver.

\subsection{Consistency Model}
\label{sec:Consistency_Model}

\begin{figure}[tbp]
\centering
\includegraphics[width=0.8\textwidth]{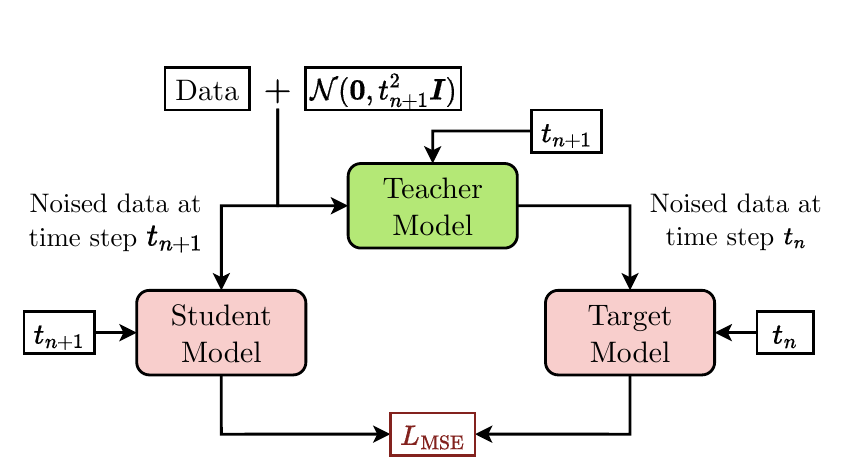}
\caption{
Illustration of the consistency distillation process distilling the diffusion model of \ccedm (teacher model) into a consistency model (student and target model). The student model is updated via gradient descent and the target model is updated as an exponential moving average of the student model weights.
}
\label{fig:consistency_distillation}
\end{figure}

Consistency Models (CM)~\cite{song2023consistency} are a recently introduced generative architecture. 
They allow for single-step or multi-step generation with the same model and can be trained standalone or distilled from a diffusion model that has already been trained.
A \textit{consistency model} $\boldsymbol{f}_\Phi$ with weights $\Phi$ is trained to estimate the \textit{consistency function} $\boldsymbol{f}$ from data. 
The consistency function is defined as $\boldsymbol{f}: (\mathbf{x}_t, t) \rightarrow \mathbf{x}_\epsilon$ and is \textit{self-consistent} in the sense that any pair of $(\mathbf{x}_t, t)$ belong to the same probability flow ODE trajectory.
This means that the result of a function evaluation at any point on this trajectory leads to the same result, i.e. $\boldsymbol{f}(\mathbf{x}_t, t) = \boldsymbol{f}(\mathbf{x}_{t'}, t')$ for all $t,t' \in [\epsilon, T]$. The time interval describes the minimum noise at time step $\epsilon$ and the maximum noise at time $T$.

For sampling from a trained consistency model in a single model pass, one initializes $\mathbf{\hat{x}}_T \sim \mathcal{N}(\mathbf{0}, T^2 \boldsymbol{I})$ and performs one function evaluation to get $\hat{\mathbf{x}}_\epsilon = \boldsymbol{f}_\Phi(\mathbf{x}_T, T)$.
It is also possible to sample with multiple model passes by first evaluating $\boldsymbol{f}_\Phi(\mathbf{x}_T, T)$, and then adding noise again from $\mathcal{N}(\mathbf{0}, t^2 \boldsymbol{I})$ to denoise a second time. 
This can be done in an alternating fashion for an arbitrary number of steps. 
Often multi-step generation appears to improve sample fidelity~\cite{karras2022elucidating_EDM_diffusion, leigh2023pcdroid}, however we are able to achieve comparable fidelity to the original diffusion model with only a single model evaluation and therefore limit ourselves to this most efficient scenario.

In line with Ref.~\cite{song2023consistency}, we found improved training fidelity when distilling the consistency model from a diffusion model instead of training it individually.
For this purpose we distill the consistency model $\boldsymbol{f}_\Phi(\mathbf{x}, t)$ from the diffusion model $\boldsymbol{s}_\phi(\mathbf{x},t)$ based on the PointWise Net of \ccedm introduced in the previous Sec.~\ref{sec:Diffusion_Model}.
The distillation is performed by separating the continuous time space $[\epsilon, T]$ into $N-1$ sub intervals (we use $N=18$). 
The interval boundaries are determined by the same step size parameterisation as in the diffusion model sampling formulation~\cite{karras2022elucidating_EDM_diffusion}.
During training a random boundary time step $t_{n \in [1,N]}$ is chosen to perform the distillation.
We refer to the original diffusion model here as the \textit{teacher} model $\boldsymbol{s}_\phi(\mathbf{x}, t)$ and to the distilled consistency model during distillation as the \textit{student} model $\boldsymbol{f}_\Phi(\mathbf{x}, t)$. 
Additionally, we call the final distilled consistency model the \textit{target} model $\boldsymbol{f}_{\Phi^-}(\mathbf{x}, t)$. 
We use the self-consistency property of the consistency model for training since it requires a well trained model to obey $\boldsymbol{f}_\Phi(\mathbf{x}_{t_{n+1}}, t_{n+1}) = \boldsymbol{f}_\Phi(\mathbf{x}_{t_n}, t_{n})$.
The neighboring points $(\mathbf{x}_{t_{n+1}}, \mathbf{x}_{t_{n}})$ on the probability flow ODE trajectory are obtained by sampling $\mathbf{x} \sim p_\text{data}$, adding noise to it to get 
$\mathbf{x}_{t_{n+1}} \sim \mathcal{N}(\mathbf{x}, t_{n+1}^2 \boldsymbol{I})$ 
and performing one ODE solver step with the teacher diffusion model to compute $\mathbf{x}_{t_{n}} = \boldsymbol{s}_\phi(\mathbf{x}_{t_{n+1}}, t_{n+1})$.
This allows the student consistency model $\boldsymbol{f}_\Phi(\mathbf{x}, t)$ to be trained with the loss:
\begin{equation}
    \mathbb{E}_{t,\mathbf{x}_t,\mathbf{x}_0} \left[ \| 
    \boldsymbol{f}_\Phi(\mathbf{x}_{t_{n+1}}, t_{n+1}) 
    - 
    \boldsymbol{f}_{\Phi^-}(\mathbf{x}_{t_{n}}, t_{n})   
    \|_2^2 \right]
\end{equation}
The target model $\boldsymbol{f}_{\Phi^-}(\mathbf{x}_t, t)$ weights $\Phi^-$ are updated after every iteration as a running average of the student model weights $\Phi$. 
An overview of the distillation procedure is illustrated in Fig.~\ref{fig:consistency_distillation}.

\subsection{Training and Sampling}
\label{sec:training_and_sampling}

The diffusion model in \ccedm was trained for 2M iterations with a batch size of 128 using the Adam optimizer~\cite{adam} with a fixed learning rate of $10^{-4}$. As the final model, we use an exponential moving average (EMA) of the model weights.
We scan several values for the number of ODE solver steps $N$ and find $N=13$ optimal as with fewer steps than this, the generative fidelity 
as probed by the correct learning of physically relevant shower shapes with \ccedm deteriorates. 
This results in $2N-1$ diffusion model evaluations since the last step of the Heun ODE solver does not perform a 2nd order correction. Compared to \caloclouds with 100 function evaluations this already hints at  a significant computational speed-up.

The diffusion model used in \ccedm was distilled into a consistency model for \cccm
by using the Adam optimizer with a fixed learning rate of $10^{-4}$ for 1M iterations with a batch size of 256. 
Notably, only a single training is necessary for distilling a model which is able to perform single step generation, as opposed to the multiple trainings required for e.g. progressive distillation~\cite{salimans2022progressive, mikuni2023fast_pointcloudDiffussion, mikuni2023caloscore_v2}.

\section{Results}
\label{sec:Results}

In the following, we compare the original \caloclouds model with the improved \ccedm model and its distilled variant \cccm. 
To achieve a fair comparison between the three models, we use the same training of the Shower Flow and the same calibration procedure for all three models. 
Hence, the Shower Flow from the \ccedm model was also used for generating samples with the \caloclouds model --- a slight modification compared to the original \caloclouds model in Ref.~\cite{CaloClouds}.
This means that the samples generated with the \caloclouds model also include the energy per layer and center of gravity in $X$ and $Y$ calibration. 
For the Latent Flow and the PointWise Net of \caloclouds the same model weights as in Ref.~\cite{CaloClouds} were used.

We first show the performance of our generative models based on the same observables as discussed in Ref.~\cite{CaloClouds} in Sec.~\ref{sec:Results_Physics}.
Next, in Sec.~\ref{sec:Results_Evaluation}, we quantify the performance of the models with multiple Wasserstein-distance-based scores for the usual set of calorimeter shower observables and in Sec.~\ref{sec:Results_Classifier} we use a classifier to distinguish between simulated \geant showers and generated showers based on the calculated shower observables.
Finally in Sec.~\ref{sec:Results_Timings} we benchmark the computational efficiency of our models and compare them to the baseline simulation timing with \geant.

\subsection{Physics Performance}
\label{sec:Results_Physics}

\begin{figure*}[h]
    \centering
    \includegraphics[width=0.32\textwidth]{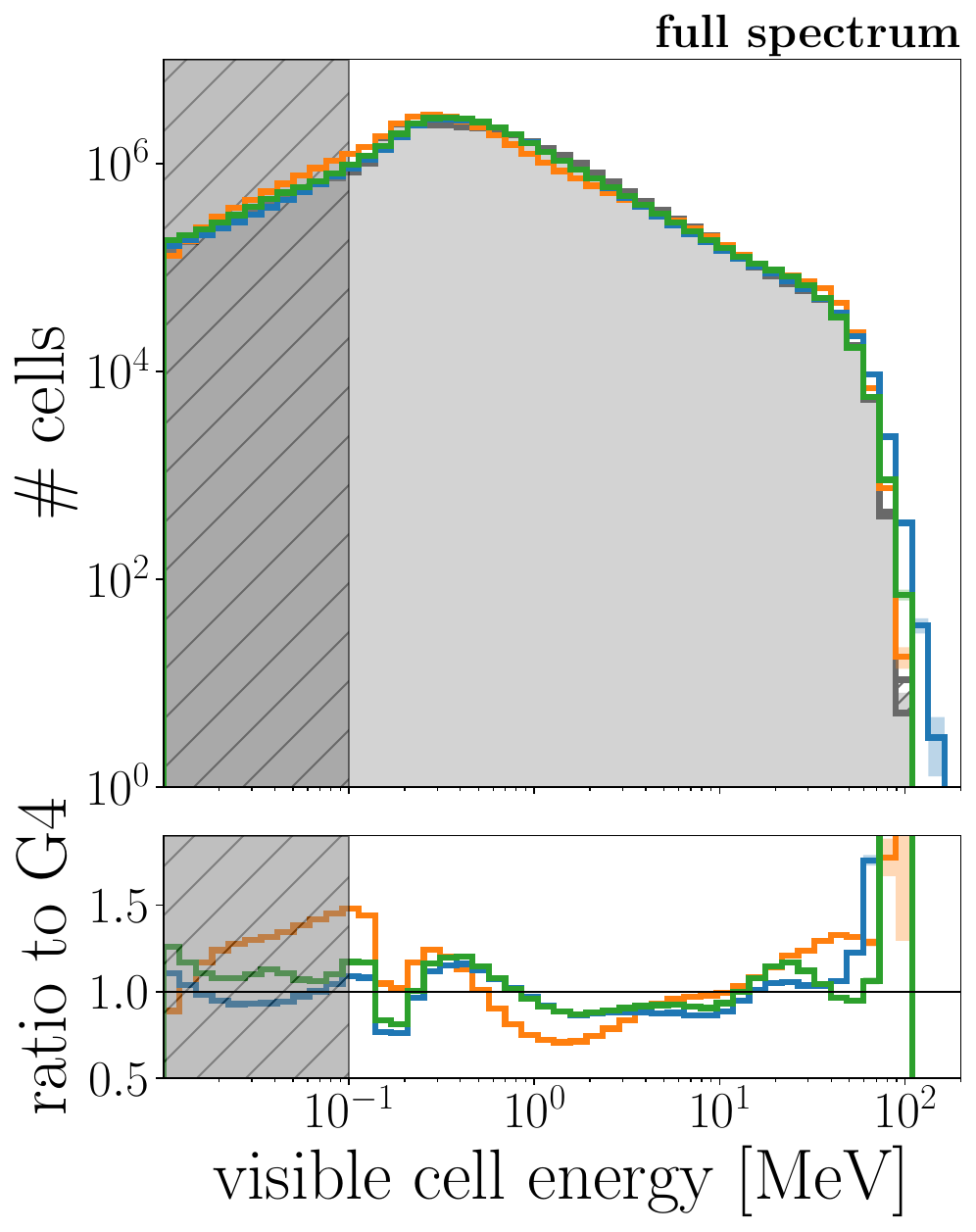}
    \includegraphics[width=0.33\textwidth]{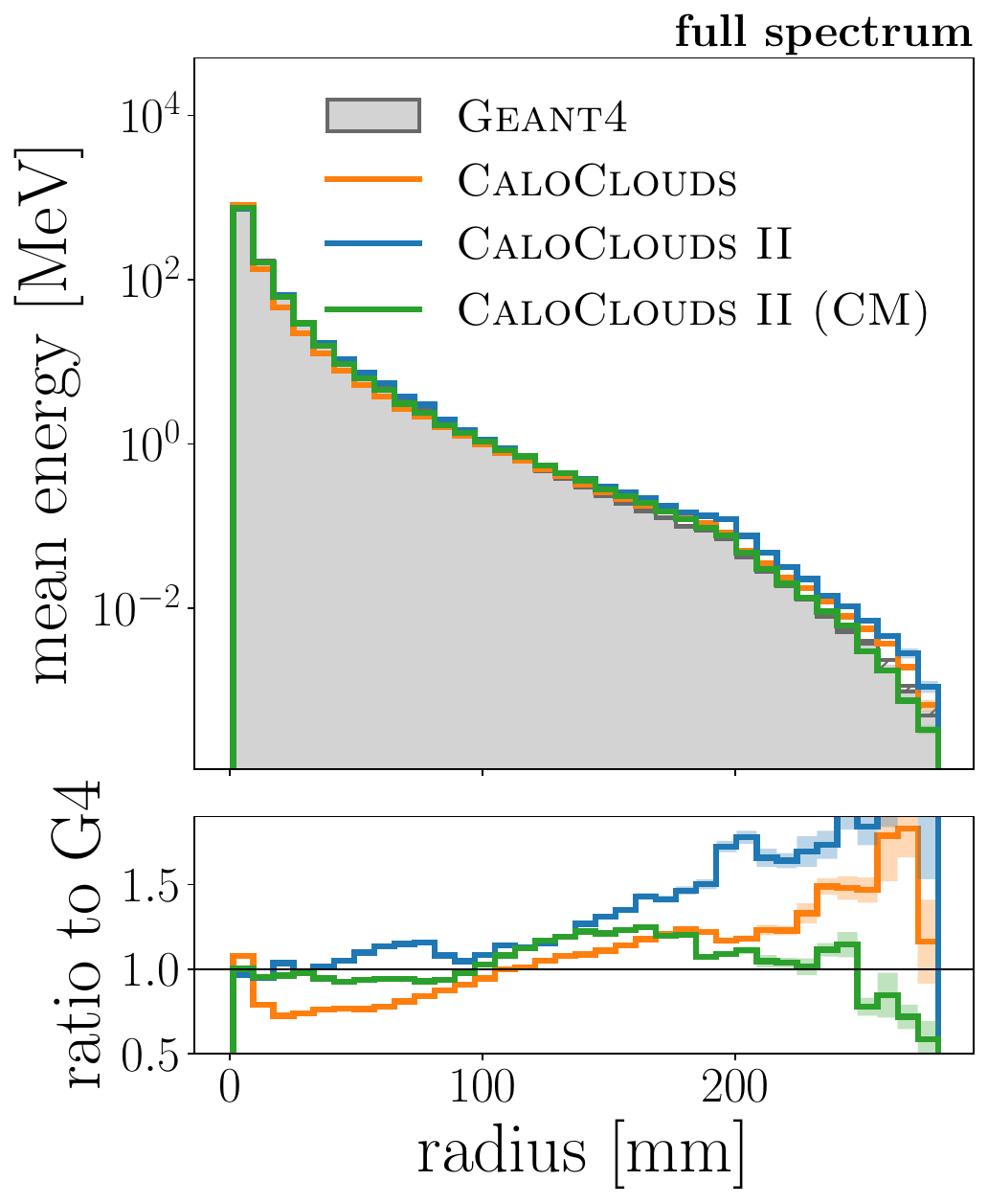}
    \includegraphics[width=0.32\textwidth]{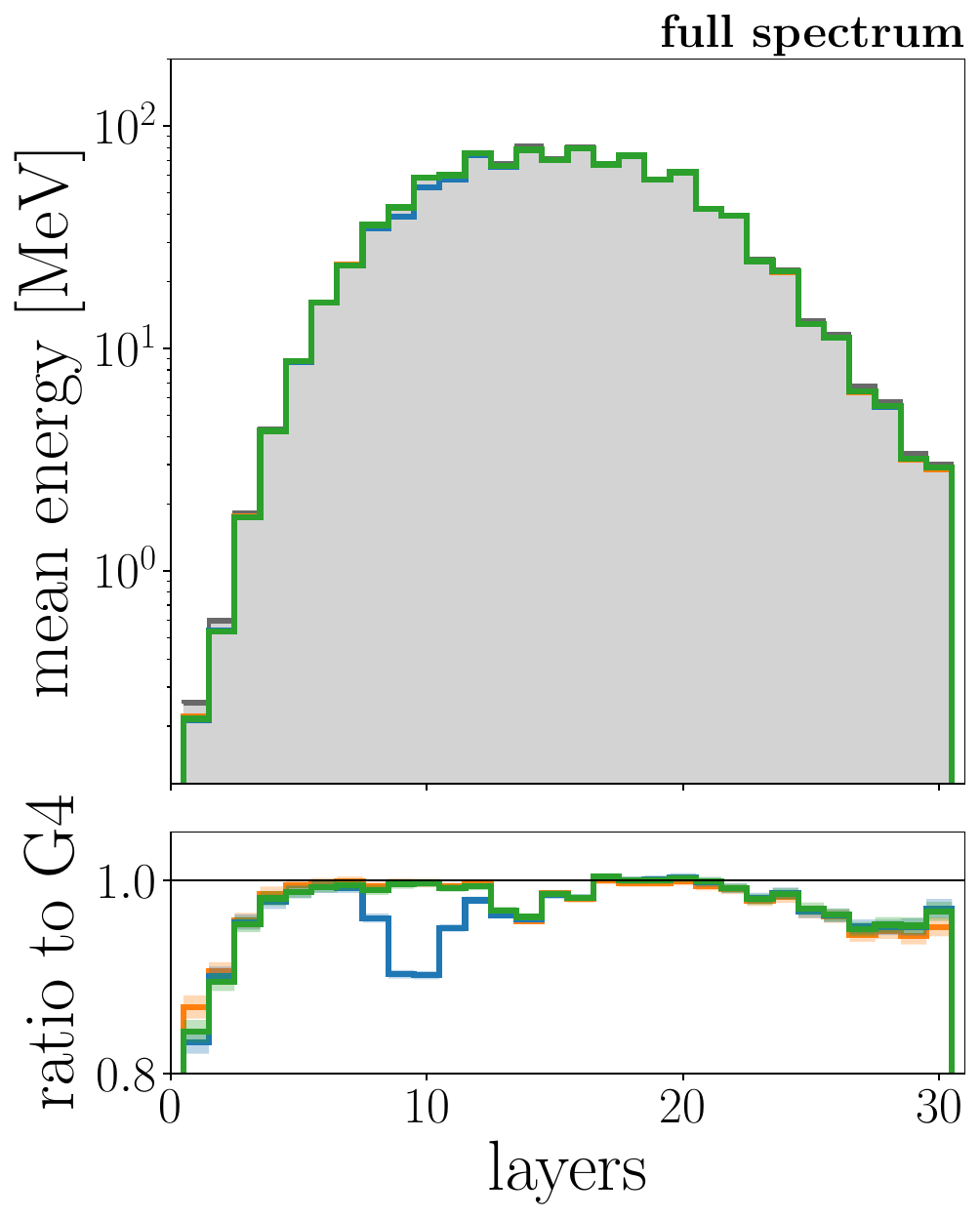}
    \caption{Histogram of the cell energies (left), radial shower profile (center), and longitudinal shower profile (right) for \geant, \caloclouds, \ccedm, and \cccm. %
    In the cell energy distribution, the region below 0.1~MeV is grayed out (see main text for details). All distributions are calculated with 40,000 events sampled with a uniform distribution of incident particle energies between 10 and 90~GeV.
    The bottom panel provides the ratio to \geant.
    The error band corresponds to the statistical uncertainty in each bin.
    }
    \label{fig:Ehits_Radial_Spinal}
    \hspace{0.5cm}
\end{figure*}

In this Section, we compare various calorimeter shower distributions from Ref.~\cite{CaloClouds} between the \geant test set and datasets generated using \caloclouds, \ccedm, and \cccm.
First, we compare various cell-level and shower observables calculated from the model generated showers 
to \geant simulations with samples of incident photons with energies uniformly distributed between 10 and 90~GeV (also referred to as \textit{full spectrum}).
In Fig.~\ref{fig:Ehits_Radial_Spinal} we investigate three representations of the energy distributed in the calorimeter cells, namely the per-cell energy distribution (left), the radial shower profile (center) and the longitudinal shower profile (right).
The per-cell energy distribution contains the energy of the cells of all showers in the test dataset.
The peak of the distribution at about 0.2~MeV corresponds to the 
most probable energy deposition of a minimum ionising particle (MIP) in the silicon sensor. 
For downstream analyses a cell energy cut at half a MIP is applied, since below this threshold the sensor response is indistinguishable from electronic noise. 
Hence this cut was applied to all showers when calculating the shower observables and scores in this section. 
All models describe the cell energy distribution reasonably well. 
For most of the range the \ccedm models perform better than \caloclouds, however there are a few outliers with energies which are too high produced by \ccedm compared to the other two models.

The radial shower profile describes the average radial energy distribution around the central shower axis (in $Z$-direction) of the ECAL.
Below a radius of about 180~mm, the distribution is well described by all three models, above 180~mm, the models deviate from \geant. Overall the \cccm model represents the \geant distribution most closely. 
Note that this is a distribution that is not directly impacted by any of the post-diffusion calibrations performed and is therefore a good benchmark for the effectiveness of the point cloud diffusion approach alone.

The longitudinal shower profile describes how much energy is deposited on average in each of the 30 calorimeter layers. 
In the previous iteration of \caloclouds it was not well modeled, but since we now calibrate the energy per layer with the improved Shower Flow for the generated point clouds it is well modelled.
However, we observe deviations in the first few layers for all three models. Since they share the same Shower Flow, we expect future improvements in this model to translate to an improved longitudinal profile.
Further, a small outlier can be seen for the \ccedm model around the $10^{th}$ layer.
The alternating higher and lower energy depositions per layer are due 
to the fact that for technical reasons, pairs of silicon sensors surrounding one tungsten absorber layer and facing opposite directions are installed into a tungsten structure with every other absorber layer. 
This results in the observed pair-wise difference in the sampling fraction between consecutive layers.

\begin{figure*}[h]
    \centering
    \includegraphics[width=1\textwidth]{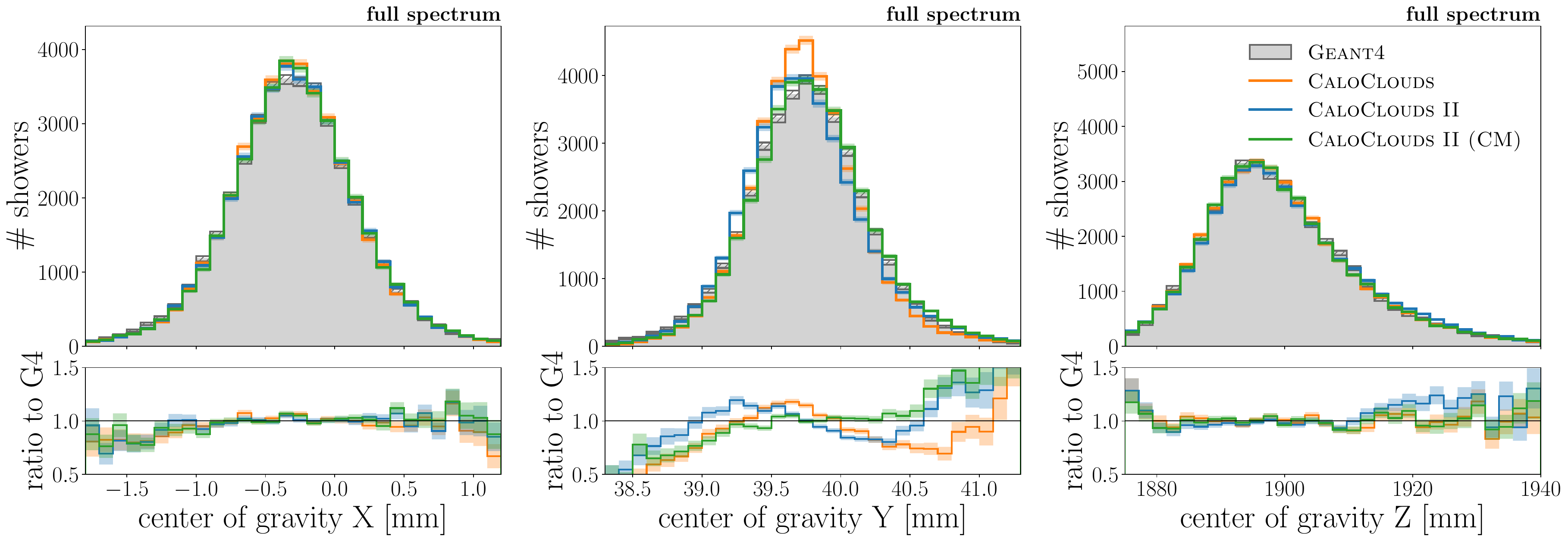}
    \caption{
    Position of the center of gravity of showers along the $X$ (left), $Y$ (center), and $Z$ (right) directions. All distributions are calculated for 40,000 showers with a uniform distribution of incident particle energies between 10 and 90~GeV.
    The error band corresponds to the statistical uncertainty in each bin.
    }
    \label{fig:CoG}
    \hspace{0.5cm}
\end{figure*}

In Fig.~\ref{fig:CoG} we show the center of gravity distribution $m_{1,i \in \{X,Y,Z\}}$ (the energy weighted shower centroid) in the $X$-, $Y$-, and $Z$-directions. 
Note that in the $X$- and $Y$-directions these distribution are calibrated for the original point cloud, before the cell-level observables are calculated.
While the $m_{1,X}$ distribution is very well modelled by all three generative models,
$m_{1,Y}$ is slightly shifted to lower center of gravity values for all models with the \caloclouds distribution additionally being marginally too narrow. 
The centers of gravity in $X$ and $Y$ behave slightly different as a magnetic field is simulated in the $Y$-direction and the active sensors are staggered in the $Y$-direction while they are all aligned in the $X$-direction.
Due to the number of hits and energy per layer calibrations applied, the distribution of $m_{1,Z}$ is very well modelled. Only in the region around 1925\,mm is the \ccedm model slightly worse than the other two models.
Overall, the three models are reasonably close to the \geant simulation in all six observables.

\begin{figure*}[h]
    \centering
    \includegraphics[width=0.49\textwidth]{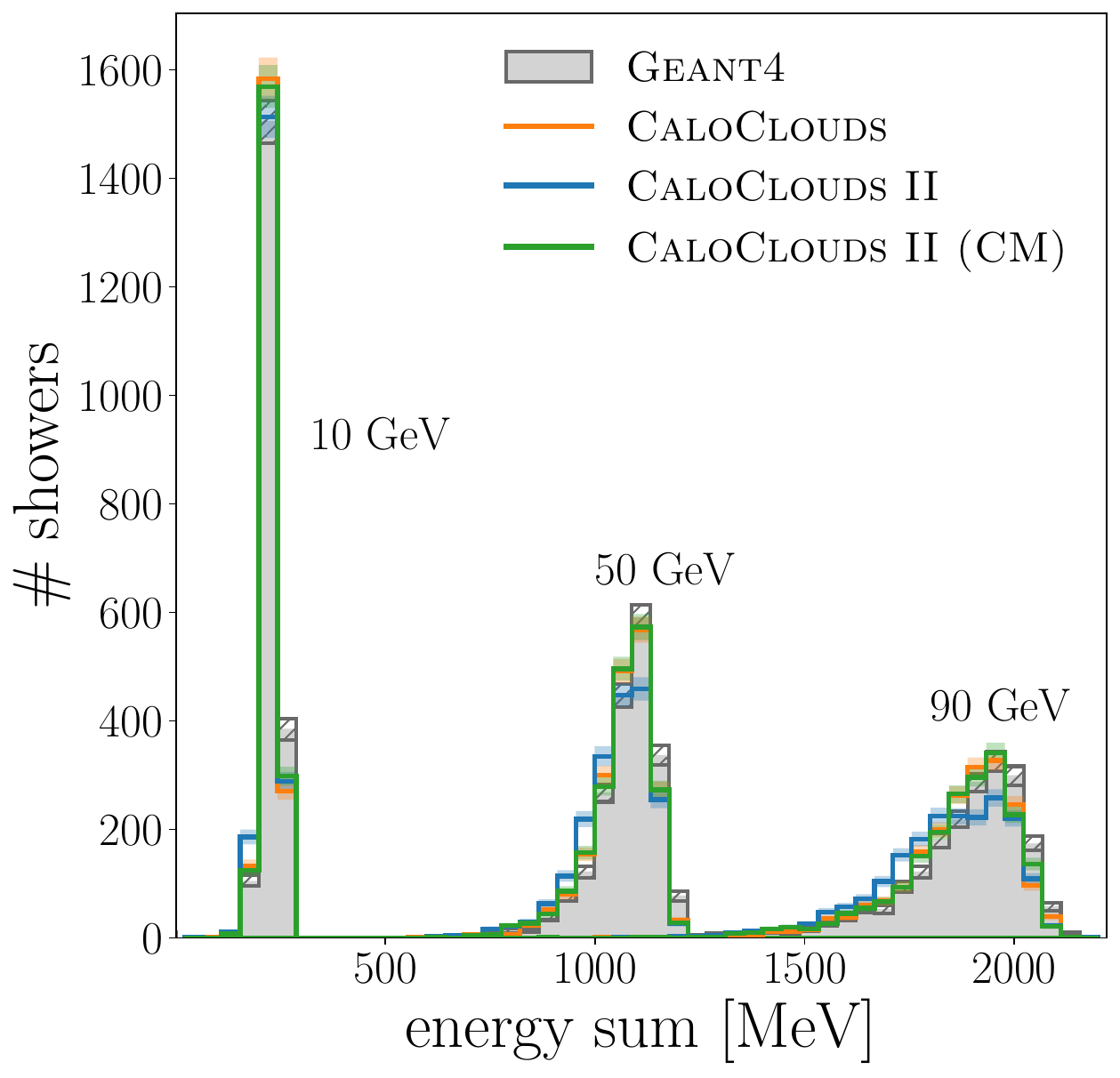}
    \includegraphics[width=0.49\textwidth]{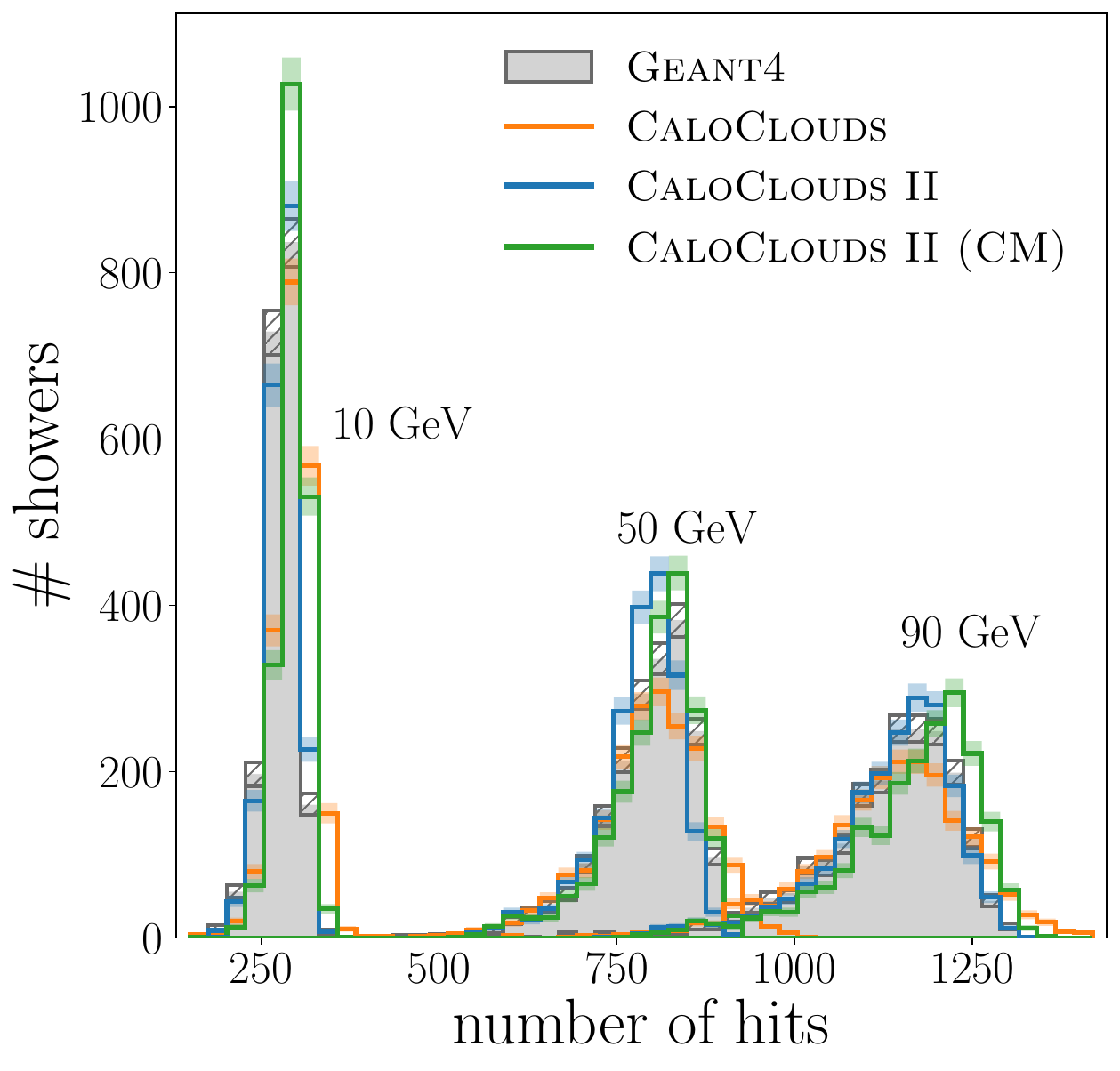}
    \caption{
    Visible energy sum (left) and the number of hits (right) distributions, for 10, 50, and 90 GeV showers.
    For each energy and model, 2,000 showers are shown.
    The error band corresponds to the statistical uncertainty in each bin.
    }
    \label{fig:Esum_Occ}
    \hspace{0.5cm}
\end{figure*}

Next, we investigate the models' fidelity for single incident photon energies of 10, 50, and 90\,GeV.
In Fig.~\ref{fig:Esum_Occ} we show the distributions of the total visible energy (left) and the distributions of the number of hits (cells with deposited energy above the half MIP threshold) for the three single energy datasets of 2k showers each.
The total energy is well represented by all three models.
The number of hits on the other hand is one of the most difficult distributions to represent well with a point cloud generative model. 
Here high fidelity is still achieved by applying the number of points calibration discussed in Sec.~\ref{sec:Model}. 
Overall the \caloclouds distributions are slightly too wide as was observed already in Ref.~\cite{CaloClouds}.
In comparison, \ccedm represents the shape of the distribution better, yet in particular for 10 and 90~GeV showers the mean is a bit too large for the \cccm generated events.
This is explainable due to the nature of the polynomial fit used for the number of points calibration. The fit does not perform very well at the edges of the incident energy space. 
It is known that extrapolation is difficult for generative models, therefore we conjecture that with a training set including lower and higher energies, the fidelity at 10 and 90\,GeV would approach the performance at 50\,GeV.
Overall the \ccedm models perform very well and are comparable in fidelity to the \caloclouds model.

\subsection{Evaluation Scores}
\label{sec:Results_Evaluation}

\begin{table*}[h]
\sisetup{
separate-uncertainty=true,
table-format=4.3(5)
}
\centering
\vspace{15pt}
\resizebox{\textwidth}{!}{
\begin{tabular}{lcccccccc}
\toprule
Simulator & $W_1^{N_\mathrm{hits}}$ & $W_1^{E_\mathrm{vis}/E_\mathrm{inc}}$ & $W_1^{E_\mathrm{cell}}$ & $W_1^{E_\mathrm{long}}$ & $W_1^{E_\mathrm{radial}}$ & $W_1^{m_{1,X}}$ & $W_1^{m_{1,Y}}$ & $W_1^{m_{1,Z}}$ \\
          & $(\times 10^{-3})$      & $(\times 10^{-3})$             & $(\times 10^{-3})$      & $(\times 10^{-3})$       & $(\times 10^{-3})$        & $(\times 10^{-3})$     & $(\times 10^{-3})$     & $(\times 10^{-3})$     \\
\midrule
\geant      & 0.7 $\pm$ 0.2         & 0.8 $\pm$ 0.2                  & 0.9 $\pm$ 0.4           & 0.7 $\pm$ 0.8            & 0.7 $\pm$ 0.1            & 0.9 $\pm$ 0.1          & 1.1 $\pm$ 0.3          & 0.9 $\pm$ 0.3   \\
            & & &\\
\caloclouds & \textbf{2.5 $\pm$ 0.3}& 11.4 $\pm$ 0.4                 & 15.9 $\pm$ 0.7          & \textbf{2.0 $\pm$ 1.3}   & 38.8 $\pm$ 1.4            & 4.0 $\pm$ 0.4          & 8.7 $\pm$ 0.3          & 1.4 $\pm$ 0.5 \\
\ccedm      & 3.6 $\pm$ 0.5         & 26.4 $\pm$ 0.4                 & \textbf{15.3 $\pm$ 0.6} & 3.7 $\pm$ 1.6            & 11.6 $\pm$ 1.5            & \textbf{2.4 $\pm$ 0.4} & \textbf{7.6 $\pm$ 0.2} & 3.9 $\pm$ 0.4          \\
\cccm       & 6.1 $\pm$ 0.7         & \textbf{9.8 $\pm$ 0.5}         & 16.0 $\pm$ 0.7          & \textbf{2.0 $\pm$ 1.4}   & \textbf{8.3 $\pm$ 1.9}    & 3.0 $\pm$ 0.4          & 9.5 $\pm$ 0.6          & \textbf{1.2 $\pm$ 0.5} \\
\bottomrule 
\end{tabular}
}
\vspace{15pt}
\caption{Model performance comparison with 1-Wasserstein distance based scores for various standardized shower observables. The values presented are the mean and standard deviation of 10 calculated scores comparing 50k \geant and 50k~generated showers.
}
\label{table:wasserstein_scores}
\end{table*}

We next investigate the performance of all three \caloclouds models by calculating scores from the high level calorimeter shower observables considered in the previous Section. 
This allows us to put a number on the fidelity observed in plots presented in the previous Sec.~\ref{sec:Results_Evaluation} and not only rely on comparing distributions by eye.

The following observables are considered in order to calculate the one-dimensional scores:
The number of hits (cells with energy depositions above the half MIP threshold) $N_\mathrm{hits}$, the sampling fraction (the ratio of the visible energy deposited in the calorimeter to the incident photon energy) $E_\mathrm{vis}/E_\mathrm{inc}$, the cell energy $E_\mathrm{cell}$%
, the center of gravity in the $X$-, $Y$-, and $Z$-directions $m_{1,i \in \{X,Y,Z\}}$, and ten observables each for the longitudinal energy $E_{\mathrm{long},i \in [1,10]}$ and for the radial energy $E_{\mathrm{radial},i \in [1,10]}$.
The ten observables for the longitudinal (radial) energy depositions are computed with the energy clustered in consecutive layers (concentric regions) such that on average all 10 observables $E_{\mathrm{long},i \in [1,10]}$ and $E_{\mathrm{radial},i \in [1,10]}$ are computed with the same statistics. 
Further details on these in total 20 observables can be found in Appendix~\ref{app:radial_long_energy}.

To compare the distributions of these observables between \geant and the three generative models, we calculate the 1-Wasserstein distance $W_1$ --- also known as the earth movers distance --- between each pair of distributions. 
The advantages of the Wasserstein distance are that it is an unbinned estimator, for one-dimensional distributions it is computationally efficient to calculate, and no  hyperparameter choices have to be made apart from the number of events used for comparison.

Following earlier works using Wasserstein distance based model evaluation scores to compare generative models~\cite{kansal2022particle,Kansal_2023}
, we calculate the distance between observables calculated from 50k \geant and 50k model generated showers.
This is done $10\times$ for independent uniformly distributed samples and we report the mean and standard deviation of the scores in Tab.~\ref{table:wasserstein_scores}. 
For this purpose, we simulated 500k \geant samples and generated 500k showers with each \caloclouds model.
To have all scores in a similar order of magnitude, we standardize each observable before we calculate the $W_1$ score.
For the layer energy and radial energy scores, $W_1^{E_\mathrm{long}}$ and $W_1^{E_\mathrm{radial}}$, we report the average Wasserstein distance over all ten bins.
The hit energy score $W_1^{E_\mathrm{cell}}$ is calculated for 50k cell hits. 
In addition to the generative model scores, we also calculate the scores for \geant itself, comparing 50k \geant showers to a separate set of 50k \geant showers.

As can be seen in Tab.~\ref{table:wasserstein_scores}, 
most model scores are quite close together.
We observe a few outliers, i.e. in the sampling fraction score $W_1^{E_\mathrm{vis}/E_\mathrm{inc}}$ the \caloclouds and \ccedm models are much better \ccedm model and in the radial energy score $W_1^{E_\mathrm{radial}}$ the \ccedm models outperform \caloclouds, which is in line with the histogram shown in Fig.~\ref{fig:Ehits_Radial_Spinal}.
Overall, \cccm appears to produce higher fidelity showers than the other two models, since it has the best score in four of the scores and does not exhibit any large outliers compared to the other two models. 
However, as can also be seen in the histograms in Sec.~\ref{sec:Results_Physics}, none of the scores --- with the exception of $W_1^{m_{1,Z}}$ --- quite reaches the fidelity of the \geant truth itself.
Hence we conclude that while all three models generate high fidelity ECAL showers, they should be further improved to match \geant exactly in the future. 

As a side note, the Wasserstein distance can be heavily impacted by outliers in the distributions.
Therefore it does not always correlate well with the distribution shape observed in histograms. 
However, the scores complement the visual inspections of histograms and distributions shown in Sec.~\ref{sec:Results_Physics} well. 

While useful for comparing generative architectures, 1-Wasserstein distances only consider each dimension of the problem individually. 
Of course, a successful generative model should also accurately describe higher order correlations. We investigate this in the next Section.

\subsection{Classifier Scores}
\label{sec:Results_Classifier}

\begin{table*}[h]
\sisetup{
separate-uncertainty=true,
table-format=4.3(5)
}
\centering
\vspace{15pt}
\begin{tabular}{lc}
\toprule
Simulator & AUC  \\
\midrule
\caloclouds & 0.999 $\pm$ 0.001 \\
\ccedm      & 0.928 $\pm$ 0.001   \\
\cccm       & \textbf{0.923 $\pm$ 0.001}   \\
\bottomrule 
\end{tabular}
\vspace{15pt}
\caption{Model performance comparison with area under the receiver operating characteristic curve (AUC) score.}
\label{table:classifier-scores}
\end{table*}

We further compare the model generated showers to the \geant simulation by training a fully connected high-level classifier using the shower observables discussed in the previous Sec.~\ref{sec:Results_Evaluation} to distinguish between model generated and \geant simulated showers.
The 25 input shower observables are the ten radial and longitudinal energy observables, as well as the three center of gravity variables and the number of hits and total visible energy. %
For the datasets, we use 500k \geant showers and 500k showers generated by each generative model.
A 80\%, 10\%, 10\% data split is applied, resulting in a training set of 800k showers and a validation and test set with 100k showers each. 

The classifier is implemented as a fully connected neural network with three layers (containing 32, 16, 8 nodes respectively) 
with LeakyReLU~\cite{Maas13rectifiernonlinearities} activation functions, and one output node with a Sigmoid activation. 
It is trained with the Adam optimizer~\cite{adam} for 10 epochs for each dataset using a binary cross-entropy loss. 
The final model epoch is chosen based on the lowest validation loss.

To evaluate the classifier we use the area under the receiver operating characteristic curve (AUC) score calculated on the test set.
This kind of \textit{classifier score} is also used in other publications evaluating generative models in high energy physics such as Ref.~\cite{krause2021caloflow, krause2021caloflow2, cresswell2022caloman, krause2023caloflow_forCaloChallenge_DS1, buckley2023inductive_caloFlow, Kansal_2023, das2023understand_limitation_genModels}.
In case the classifier can perfectly separate the \geant and model generated datasets, it will result in an AUC = 1.0. 
For a generated dataset that is indistinguishable from \geant simulation, we expect a confused classifier with an AUC = 0.5.
Values in between are difficult to interpret in absolute terms, but can give a rough indication of how well the generative models are performing compared to each other.
Note that its already not trivial to implement a generative model that achieves AUC values below 1.0.

We trained the classifier ten times with a different train/ test/ validation data split each time. 
In Tab.~\ref{table:classifier-scores} we present the mean AUC and standard deviation of these ten classifier trainings.
The \caloclouds generated dataset performs the worst, leading to an almost perfect classification with AUC = 0.999. 
The two \ccedm variants both have a better score clearly separated from an AUC = 1.0.
With an AUC = 0.923 the \cccm model performs slighly better than the \ccedm model. 
For most events, both models result in a separability from the \geant simulated showers, but constitute a clear improvement over the baseline \caloclouds implementation. 
The better performance of the \ccedm variants is likely due to the improved radial energy distribution, as we observed a rather large deviation in the $W_1^{E_\mathrm{radial}}$ score and in the radial distributions in Fig.~\ref{fig:radial_bins}.

\subsection{Timing}
\label{sec:Results_Timings}

\begin{table*}[h]
\sisetup{
separate-uncertainty=true,
table-format=4.3(5)
}
\centering
\vspace{15pt}
\resizebox{\textwidth}{!}{
\begin{tabular}{llcc|cr}
\toprule
Hardware & Simulator & NFE & Batch Size & {Time / Shower [ms]} & Speed-up \\
\midrule
CPU & \geant & & & 3914.80 $\pm$ 74.09 & $\times 1$ \\
    & & &\\
    & \caloclouds & 100 & 1 & 3146.71 $\pm$ 31.66 & $\times 1.2$ \\   %
    & \ccedm & 25 & 1 & 651.68 $\pm$ 4.21 & $\times 6.0$ \\    %
    & \cccm & 1 & 1 & 84.35 $\pm$ 0.22 & $\times 46$ \\  %
    & & &\\
GPU & \caloclouds & 100 & 64 & 24.91 $\pm$ 0.72 & $\times 157$ \\  %
    & \ccedm & 25 & 64 & 6.12 $\pm$ 0.13 & $\times 640$ \\  %
    & \cccm & 1 & 64 & 2.09 $\pm$ 0.13 & $\times 1873$ \\  %
\bottomrule 
\end{tabular}
}
\vspace{15pt}
\caption{Comparison of the computational performance of \caloclouds, \ccedm, and \cccm to the baseline \geant simulator on a single core of an Intel\textsuperscript{\tiny\textregistered} Xeon\textsuperscript{\tiny\textregistered} CPU E5-2640 v4 (CPU) and on an NVIDIA\textsuperscript{\tiny\textregistered} A100 with 40~GB of memory (GPU). 2,000 showers were generated with incident energy uniformly distributed between 10 and 90~GeV. Values presented are the means and standard deviations over 10 runs. The number of function evaluations (NFE) indicate the number of diffusion model passes.
}
\label{table:timing}
\end{table*}

In this Section, we benchmark the average time to produce a single calorimeter shower with the three models considered and investigate the speed-up over the baseline \geant simulation.
The timing results are presented in Tab.~\ref{table:timing}.

On both a single CPU and on an NVIDIA\textsuperscript{\tiny\textregistered} A100 GPU we generated $25\times$ 2,000 showers with the same uniform energy distribution between 10 and 90~GeV. 
We report the mean and standard deviation of generating these showers.
In particular the timing on a single CPU is interesting for current applications of generative models in high energy physics, as CPUs are much more widely available than GPUs and the current computing infrastructure relies on simulations run on CPUs.
Further, the single CPU timing facilitates a direct comparison to the \geant simulation.
Here \caloclouds already yields a speed up of $1.2\times$, but with less sampling steps $\ccedm$ achieves a speed up of $6.0\times$. 
However, when implementing the consistency distillation, we achieve a speed up of $46\times$ with the \cccm model even surpassing previous generative models on the same kind of dataset such as the BIB-AE~\cite{gettinghigh} by about a factor 5. 

On an NVIDIA\textsuperscript{\tiny\textregistered} A100 GPU the \caloclouds model achieves a speed up of $157\times$, \ccedm achieves $640\times$, and \cccm achieves $1873\times$ speed up over the baseline \geant simulation on a single CPU.
Note that \geant is currently not compatible with GPUs and that GPUs are significantly more expensive than CPUs.

For reference, the training of the \caloclouds model on similar NVIDIA\textsuperscript{\tiny\textregistered} A100 GPU hardware took around 80 hours for 800k iterations with a batch size of 128, while training of the \ccedm model took around 50 hours for 2 million iterations with the same batch size. The consistency distillation for 1 million iterations with a batch size of 256 took about 100 hours.

The speed up between \caloclouds and \ccedm is the result of a combination of the improved diffusion paradigm requiring a reduced number of function evaluations as well as the removal of the latent flow. 
The speed up due to the consistency model in \cccm yields another large factor, since only a single model evaluation is performed.
Both models would be slightly slower when applied in conjunction with the Latent Flow of the \caloclouds model as one evaluation of the Latent Flow is about 50\% slower than a single evaluation of the PointWise Net. 
For a large number of model passes of the PointWise Net in the diffusion framework, the efficiency of the Latent Flow is negligible. 
However when we consider \cccm with a single model pass, the application of the Latent Flow would have a noticeable impact on computational performance.  
Therefore, we removed the Latent Flow in favour of model efficiency as we did not see any improvement in generative fidelity when using it in the \ccedm framework.

\section{Conclusions}
\label{sec:Conclusion}

\caloclouds was the first generative model to achieve high-fidelity highly-granular photon calorimeter showers in the form of point clouds with a number of points of $\mathcal{O}(1000)$.
Due to their sparsity, describing calorimeter showers as point clouds is computationally more efficient than 
describing them with fixed data structures, i.e. 3d images. 
Additionally, as the point clouds are based on clustered \geant steps, they allow for a translation-invariant and geometry-independent shower representation.
Such cell-geometry-independent models could be easily adapted for fast simulations of calorimeters with non-square cell geometries, i.e. hexagonal cells as used in the envisioned CMS HGCAL~\cite{CMS_hgcal}.

With \ccedm we introduce a more streamlined version of \caloclouds utilizing the advanced diffusion paradigm from Ref.~\cite{karras2022elucidating_EDM_diffusion}.
It allows for sampling with less model evaluations and for distillation into a consistency model.
Using the consistency model in \cccm, generation with a single model evaluation is possible and results in a greatly improved computational efficiency and a speed up of $46\times$ over \geant on a single CPU.
This single event CPU performance is particularly promising for introducing a generative model into  existing \geant-based simulation pipelines.
As opposed to other diffusion distillation methods like progressive distillation, consistency distillation only requires a single training to distill the diffusion model in \ccedm into a single step generative model, further emphasising the computational advantage of the models presented here.
To our knowledge, this constitutes the first application of a consistency model to calorimeter data.

We compare all three point cloud generative models using one-dimensional distributions and a classifier-based measure  and find comparable performance with a slight advantage for the \ccedm variants. In particular, the \cccm model exhibits superior performance while being significantly more computationally efficient.
It is counter-intuitive, that a distilled consistency model outperforms the original diffusion model, however, it is known that ODE solvers might introduce errors in earlier denoising steps that are then propagated to the generated samples~\cite{karras2022elucidating_EDM_diffusion}. 
The consistency model avoids this since we use it for single-shot generation.
Yet, slight deviations from the \geant simulations are still visible in various shower observables.
Further improvements could likely be achieved by investigating more complex architectures for the diffusion model such as fast transformer implementations~\cite{dao2022flashattention}, equivariant point cloud (EPiC) layers~\cite{Buhmann:2023pmh}, or cross-attention~\cite{vaswani2023attention_is_all_you_need}.

During the completion of this manuscript, another EDM diffusion based model with subsequent consistency distillation was shown to achieve good fidelity when generating particle jets in the form of point clouds with up to 150 points~\cite{leigh2023pcdroid}.
While technically a similar approach, in our case the consistency model does not lose generative fidelity compared to the diffusion model and we demonstrate the generation of two orders of magnitude more points (6000 vs 150). 

In conclusion, the \ccedm model generates high fidelity electromagnetic showers when benchmarked on various shower observables against the baseline \geant simulation.
In combination with consistency distillation the \cccm model yields an accurate simulator, which is significantly faster than \geant on identical hardware. This constitutes an important step towards the integration of point-cloud based generative models in actual simulation workflows.

\appendix
\section{Radial and longitudinal energy observables}
\label{app:radial_long_energy}

To explore the radial and longitudinal energy profile shown in Fig.~\ref{fig:Ehits_Radial_Spinal} further and to calculate the evaluation scores in Sec.~\ref{sec:Results_Evaluation}, we define ten radial and longitudinal energy observables for the calorimeter showers.

Respectively, the ten observables are defined such that energy is clustered in each observable with an equal amount of statistics.
Put differently, the energy is binned in ten quantiles with approximately the same number of cell hits in each quantile.
The energy bins are defined by the quantiles calculated on the \geant test set with 40,000 events.
While the bin edges are precisely defined for the radial energy, we round the bin edges of the longitudinal observables to the nearest layer integer number.

\begin{figure}[h]
    \centering
    \includegraphics[width=1\textwidth]{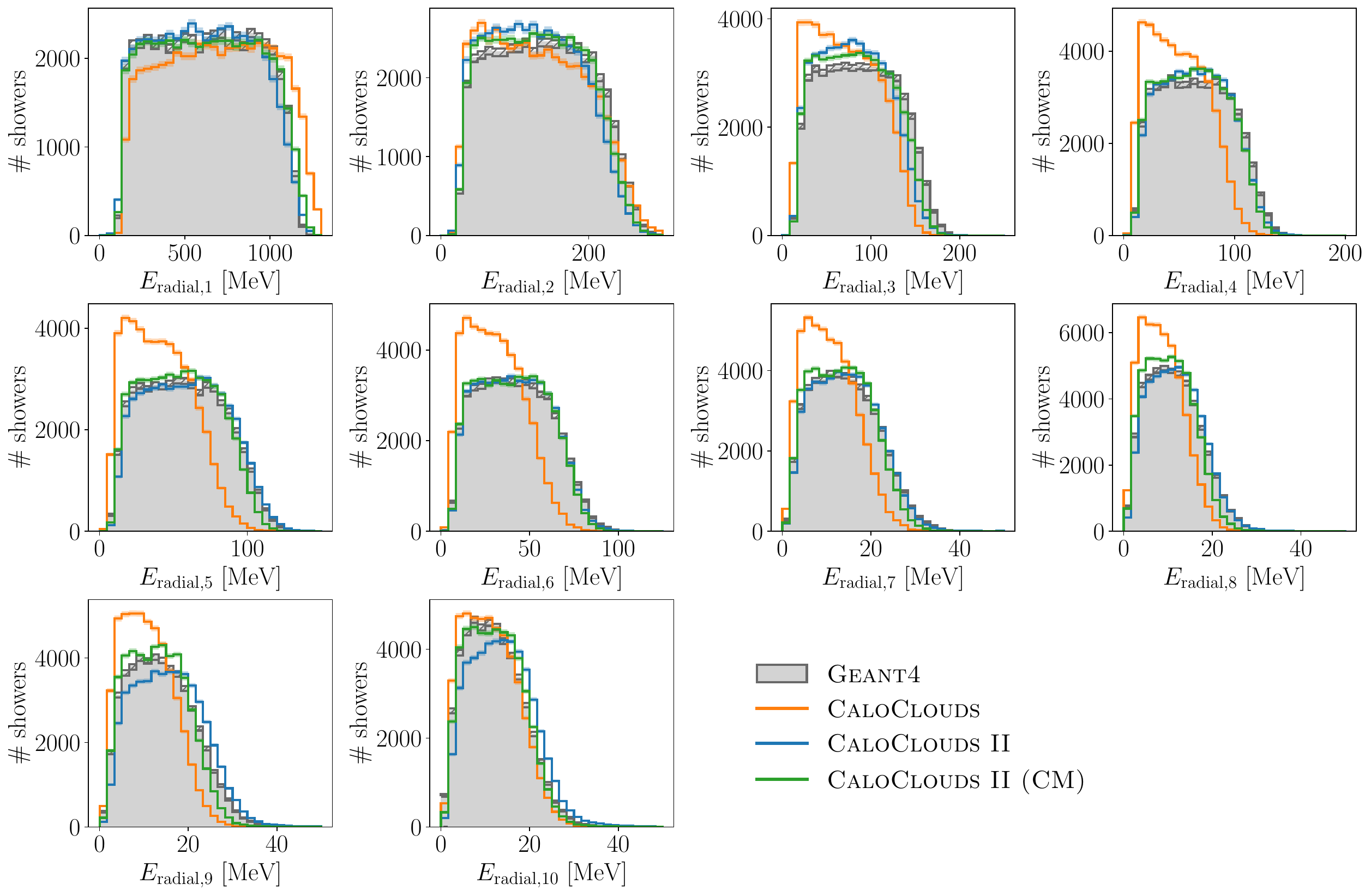}
    \caption{
    Radial energy observables for 50,000 showers. 
    The error band corresponds to the statistical uncertainty in each bin.
    }
    \label{fig:radial_bins}
\end{figure}

\begin{figure}[H]
    \centering
    \includegraphics[width=1\textwidth]{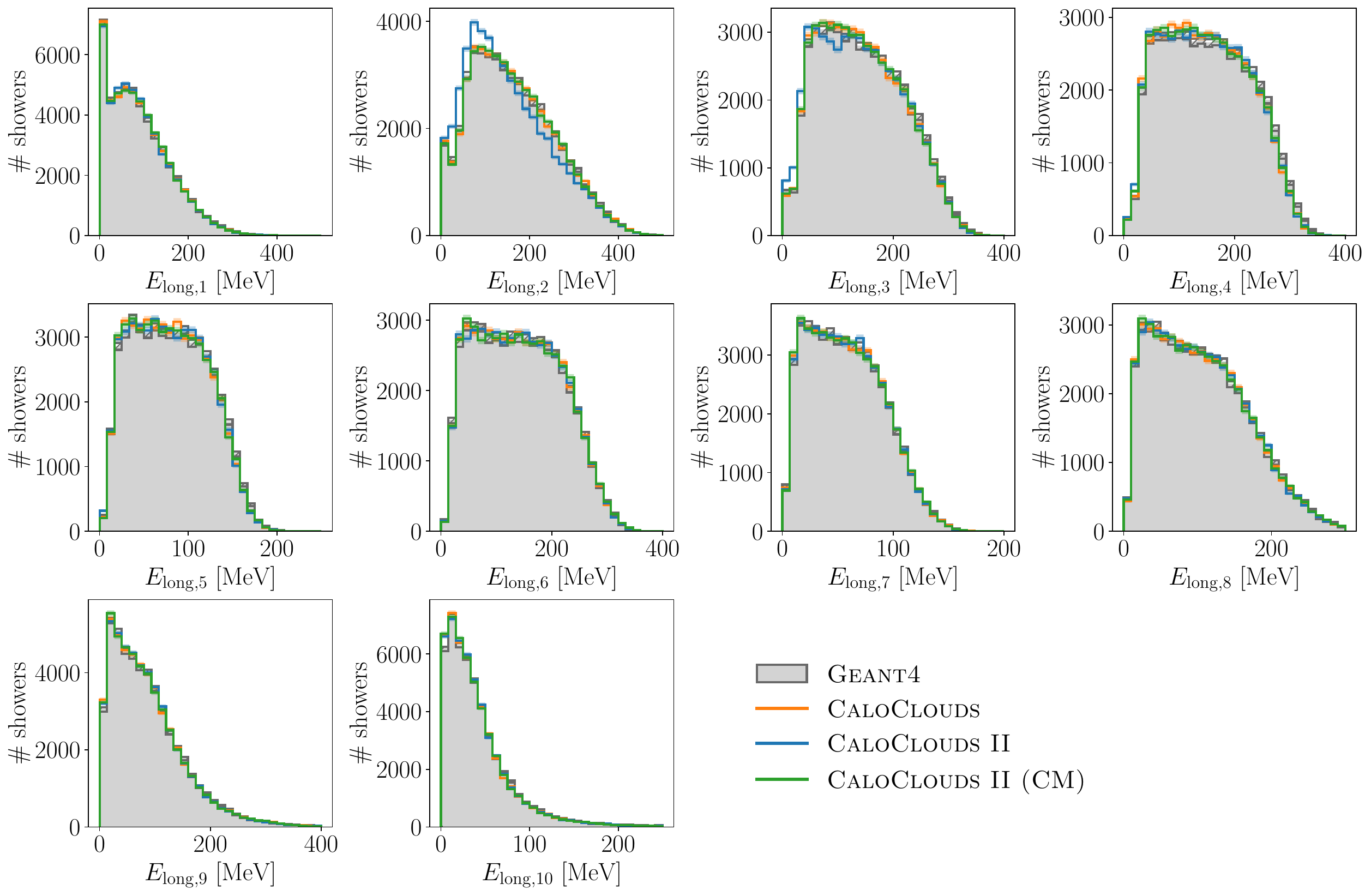}
    \caption{
    Longitudinal energy observables for 50,000 showers. 
    The error band corresponds to the statistical uncertainty in each bin.
    }
    \label{fig:layer_bins}
\end{figure}

Histograms of the radial energy observables $E_{\mathrm{radial},i \in [1,10]}$ are shown in Fig.~\ref{fig:radial_bins} and of longitudinal energy observables $E_{\mathrm{long},i \in [1,10]}$ in Fig.~\ref{fig:layer_bins}. The bin edges for all observables are given in Tab.~\ref{table:radial_long_bins}.

\begin{table*}[h!]
\sisetup{
separate-uncertainty=true,
table-format=4.3(5)
}
\centering
\vspace{15pt}
\resizebox{\textwidth}{!}{
\begin{tabular}{lrrrrrrrrrrr}
\toprule
Bin edges & 0 & 1 & 2 & 3 & 4 & 5 & 6 & 7 & 8 & 9 & 10 \\
\midrule
Edges for $E_{\mathrm{radial}, i\in [1,10]}$ [mm] & 0 & 6.6 & 9.8 & 13.0 & 17.0 & 23.4 & 33.6 & 40.1 & 48.5 & 68.8 & 300 \\
Edges for $E_{\mathrm{long}, i\in [1,10]}$ [layer] & 1 & 9 & 12 & 14 & 16 & 17 & 19 & 20 & 22 & 25 & 30  \\
\bottomrule 
\end{tabular}
}
\vspace{15pt}
\caption{Bin edges for calculating the radial and longitudinal energy observables $E_{\mathrm{radial}, i\in [1,10]}$ and  $E_{\mathrm{long}, i\in [1,10]}$. Determined for ten quantiles each including approximately the same number of cell hits. All bins are half-open, except the last bin.
}
\label{table:radial_long_bins}
\end{table*}

\acknowledgments

We thank Dirk Krücker for the valuable comments on the manuscript.
This research was supported in part by the Maxwell computational resources operated at Deutsches Elektronen-Synchrotron DESY, Hamburg, Germany. This project has received funding from the European Union’s Horizon 2020 Research and Innovation programme under Grant Agreement No 101004761. 
We acknowledge support by the Deutsche Forschungsgemeinschaft under Germany’s Excellence Strategy – EXC 2121  Quantum Universe – 390833306 
and via the KISS consortium (05D23GU4, 13D22CH5) funded by the German Federal Ministry of Education and Research BMBF in the ErUM-Data action plan.
E.B. is partially funded by a scholarship from the Friedrich Naumann Foundation for Freedom.
A.K. has received support from the Helmholtz Initiative and Networking Fund’s initiative for refugees as a refugee of the war in Ukraine.

\bibliographystyle{JHEP.bst}
\bibliography{main.bib}

\end{document}